\documentclass[epsf,twocolumn,preprintnumbers]{revtex4}
\usepackage{graphics}
\usepackage{graphicx}
\usepackage{dcolumn} 
\usepackage{bm}
\usepackage{color}
\usepackage{epsfig}
\usepackage{comment}
\usepackage{url}
\usepackage[normalem]{ulem}
\usepackage{xcolor}

\newcommand\redsout{\bgroup\markoverwith{\textcolor{red}{\rule[0.5ex]{2pt}{0.4pt}}}\ULon}

\newcommand{\beq}{\begin{equation}}
\newcommand{\eeq}{\end{equation}}
\newcommand{\bea}{\begin{eqnarray}}
\newcommand{\eea}{\end{eqnarray}}

\newcommand{\lno}{LaNiO$_2$}

\newcommand{\bnoas}{Ba$_2$NiO$_2$(AgSe)$_2$}

\usepackage{calligra}
\DeclareMathAlphabet{\mathcalligra}{T1}{calligra}{e}{f}

\begin{document}
\title{A $d^8$ anti-Hund's Singlet Insulator in an Infinite-layer Nickelate}

\author{Hyo-Sun Jin$^1$}
\author{Warren E. Pickett$^{2}$}
\email{wepickett@ucdavis.edu} 
\author{Kwan-Woo Lee$^{1,3}$}
\email{mckwan@korea.ac.kr} 
\affiliation{
 $^1$Division of Display and Semiconductor Physics, Korea University, Sejong 30019, Korea\\
 $^2$Department of Physics and Astronomy, University of California Davis, Davis CA 95616, USA\\
 $^3$Department of Applied Physics, Graduate School, Korea University, Sejong 30019, Korea
}
\date{\today}
\begin{abstract}
The status of nickelate superconductors in relation to cuprate high temperature superconductors is one of the concepts being discussed in high temperature superconductivity in correlated transition metal oxides. 
New additions to the class of infinite layer nickelates can provide essential input relating to connections or distinctions. 
A recently synthesized compound \bnoas, which contains isolated `infinite layer' NiO$_2$ planes, may lead to new insights. Our investigations have discovered that, at density functional theory mean field level, the ground state consists of an unusual $e_g$ singlet on the Ni$^{2+}$ ion 
 arising from large but separate Mott insulating gaps in both $e_g$ orbitals, 
but with different, anti-Hund's, spin directions of their moments. 
This textured singlet incorporates at the least new physics, and potentially a new platform for nickelate superconductivity, 
which might be of an unconventional form for transition metal oxides due to the unconventional undoped state. We include in this paper a comparison of electronic structure parameters of \bnoas~with a better characterized infinite layer nickelate LaNiO$_2$. 
We provide more analysis of the $d^8$ anti-Hund's singlet that emerges in this compound, and consider a minimally correlated wavefunction for this singlet in an itinerant background, and begin discussion of excitations -- real or virtual -- that may figure into new electronic phases. 
\end{abstract}
\maketitle

\noindent
\section{Introduction}
The discovery of superconductivity, after 30 years of attempts, in hole-doped `infinite layer' nickelates ${\cal R}$NiO$_2$ (${\cal R}$=lanthanide)\cite{H.Hwang2019}, has broadened the scope of high temperature superconductivity (HTS). Such nickelates have identical crystal structures and underlying $d^9$ metal atom configurations, yet nickelate superconductivity has been remarkably elusive.
 Although the achieved temperatures so far\cite{H.Hwang2020,Ariando2020}, T$_c$ up to 15 K, are not yet HTS, the connection is clear. The nickelates have the same square planar transition metal-oxygen layers, and the same underlying formal oxidation state with a single hole in the $3d$ shell: Ni$^{1+}$ rather than Cu$^{2+}$. In both systems, doping the $d^9$ configuration toward $d^8$ leads to superconductivity.\cite{H.Hwang2020,Ariando2020} Yet with these and other similarities, essential differences have been uncovered \cite{lee2004,choi_prr2020,arita2021,wen2021,review_our2021}.
 
 An early study by some of the authors of the `infinite layer' compound LaNiO$_2$ (Ni$^{1+}$, $d^9$) revealed substantial differences which emphasized that, while relations between the two ions are evident, Ni$^{1+}$ does not behave like Cu$^{2+}$ in these structures \cite{lee2004}.
 When strong correlation effects are included in the band calculation, there was a strong tendency toward the formation of peculiar on-site `singlet' spin character in a metallic background arising from spin compensation between the two $e_g$ orbitals. This behavior arises in spite of the formal spin-half (not spin $S$=0 or $S$=1) criterion of the formal oxidation state. The carriers at the Fermi level (and below -- the hole states) consisted of roughly equal $d_{x^2-y^2}$ spin up, $d_{z^2}$ spin down. This incipient singlet 
 leaves oxygen ions straying from the O$^{2-}$ configuration, forming the metallic background. In the same sort of calculation for Cu$^{2+}$, a spin-half antiferromagnetic (AFM) insulating state is obtained, consistent with experiment. 
 
 Another distinction from cuprates is that a $d^8$ nickelate Ba$_2$NiO$_2$(AgSe)$_2$ (BNOAS) has been reported \cite{bnoas_exp}, and it contains an infinite-layer NiO$_2$ layer. The formally $d^8$ cuprate LaCuO$_3$ is a conventional although metastable metal, with samples being oxygen deficient \cite{karpinen} and typically prepared under pressure. Other formally $d^8$ cuprates are rare. 
 The magnetic character of BNOAS is enigmatic. Conventionally, the $d^8$ configuration should be a non-magnetic singlet, with two holes in the $d_{x^2-y^2}$ orbital, or a triplet, with spin parallel holes in each of the $e_g$ orbitals. The former has zero (spin) susceptibility, and the latter would have a Curie-Weiss susceptibility with AFM ordering at some temperature, supposing conventional antialigned neighbors. 
 
 What was observed by Matsumoto {\it et al.} \cite{bnoas_exp} is (when zero field cooled) a constant value, on top of which is a rounded peak centered at T$^*$=130 K with half width $\sim$10 K -- apparently some magnetic reconstruction in an otherwise magnetically inert state. When field cooled, the susceptibility is higher below T$^*$, and a weak peak is seen at ($\frac{1}{2},\frac{1}{2}$,0) in polarized neutron scattering at 5 K, suggesting antialigned neighboring moments.  Matsumoto {\it et al.} suggested a canted AFM state of $S$=1 spins to account for the field cooling dependence below T$^*$, but other aspects of the susceptibility remain to be explained. These are some of the questions we address.   
 
 The  report of BNOAS has reinvigorated study of the two dimensional (2D) Kondo necklace model \cite{bnoas_ours}, the 1D version of which was proposed by Doniach \cite{Doniach}. The 2D model consists of a square lattice of two-orbital sites, with in-plane coupling $J$ between neighboring orbital$_1$ states and on-site Kondo coupling $K$ between orbital$_1$ and orbital$_2$. Wu {\it et al.} have applied the iPEPS (infinite projected entangled pair state) formalism to obtain the phase diagram of the anisotropic XXZ version \cite{tu_bnoas}, finding spin-singlet, supersolid, and Bose-Einstein condensed phases for various ratios $K/J$. Finite temperature studies by Singh \cite{RRPSingh} included the local $S=1$ excited states that are exchange coupled to neighboring $S=1$ states. This spin model leads to a ground state first-order transition between singlet and antiferromagnetic phases, with finite temperature behavior being amenable to cluster expansion techniques.  Separately, Kitamine {\it et al.} have pursued design of $d^8$ nickelate materials \cite{kitamine2020}, focusing on $A_2$NiO$_2X_2$ compounds with $A$ being a divalent alkaline earth cation and anions $X$=H, F, Cl, Br, I, the aim being to isolate the $d_{x^2-y^2}$ orbital and leave other $d$ orbitals just below the Fermi energy rather than to engineer magnetic behavior.

The organization of the paper is as follows. Section II provides the input to the theoretical work: crystal structure and computational methods. The non-magnetic electronic structure, including a comparison of $d^8$ and $d^9$ ions in infinite layer environments, is provided in Sec. III, and the magnetic calculations are introduced. In Sec. IV effects of correlations are introduced, and the peculiar band singlet that arises is discussed and its origin analyzed. Then some analysis of this anti-Hund's singlet and a corresponding correlated Hartree-Fock wavefunction is given in Secs. V and VI, with the intent that it might shed light on the peculiar observed magnetic behavior of BNOAS. A Discussion and Summary is provided in Sec. VII.

\begin{figure}[tbp]
  \includegraphics[width=\columnwidth]{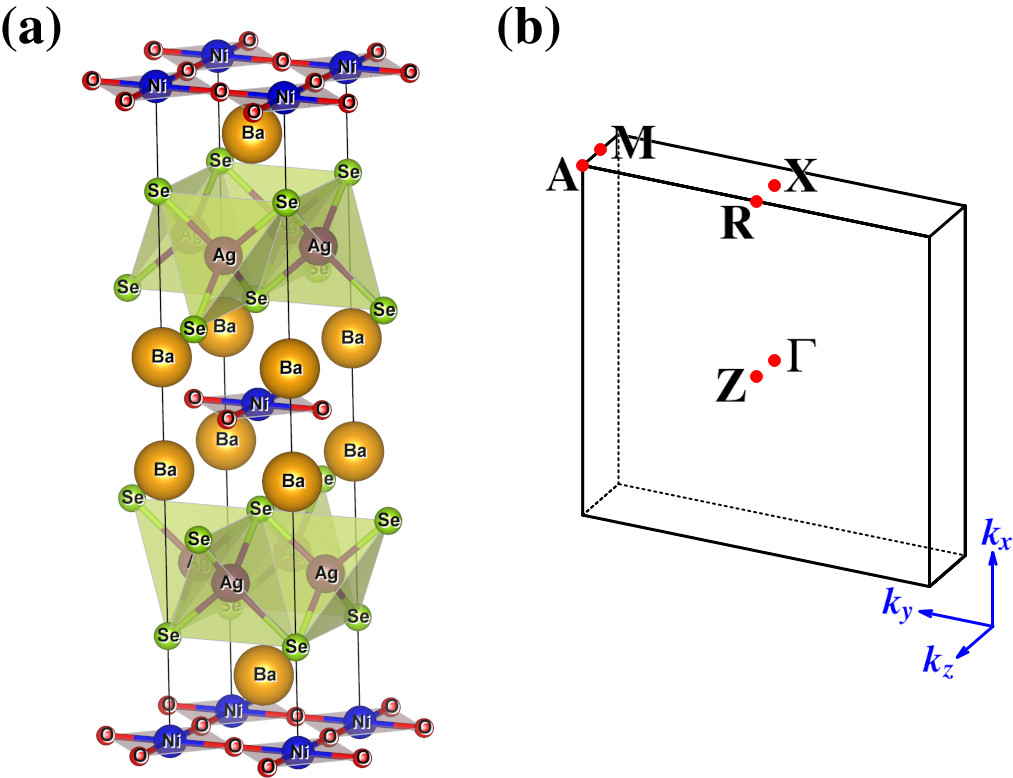}
  \caption{(a) Body-centered tetragonal structure of \bnoas.
  (b) Tetragonal Brillouin zone doubled along the $\hat{c}$-axis and high symmetry points given in the band structure plots below.
  }
  \label{str}
\end{figure}

\section{Crystal Structure and Methods} 

{\it BNOAS crystal structure.}   BNOAS is usefully considered as [Ba(AgSe)$_2$Ba]NiO$_2$ -- an active
$\infty$-layer Ni$^{2+}$O$_2$ plane separated by thick Ba(AgSe)$_2$Ba blocking layers \cite{bnoas_exp}.  
The space group is body-centered tetragonal $I4/mmm$ (\#139) \cite{bnoas_exp}, and the Ni ion has
$4/mmm$ symmetry. As displayed in Fig. \ref{str}(a), the (AgSe)$_2$ blocking layer has the FeSe structure: Ag in the central layer is coordinated tetrahedrally by Se atoms, which in the next layers are neighbored by a plane of Ba$^{2+}$ ions. Ni has no apical oxygen neighbors. 

The compound is apparently metastable: it was synthesized under 7 GPa pressure at 850 $^\circ$C. This is reminiscent of the superconducting `infinite layer' nickelates, which are only superconducting when grown in strained perovskite form using layer-by-layer synthesis techniques \cite{H.Hwang2019}, with O later removed topotactically. In BNOAS, strain supplied by the (AgSe)$_2$ layer leads to an anomalously large Ni-O separation of 2.105 \AA, which Matsumoto {\it et al.}\cite{bnoas_exp} argued should put BNOAS beyond an $S=0$ to $S=1$ transition at 2.05 \AA. Our earlier results also indicated a change in electronic behavior \cite{bnoas_ours} around 2.05 \AA. 

{\it  Theoretical methods.} Our calculations are based at the density functional theory (DFT) level using the generalized gradient approximation (GGA) \cite{gga} implemented in {\sc wien2k} \cite{wien2k}. In {\sc wien2k}, we used the same input parameters in our previous studies of BNOAS \cite{bnoas_ours}.
The in-plane lattice constant is $a$=3.96 (4.21) \AA~ for \lno~ (BNOAS). The latter anomalously large value is believed to be key to the unusual behavior of the Ni ion \cite{bnoas_exp,bnoas_ours}.

Correlation effects were included as conventionally done using the DFT+U approach \cite{ylvisaker2010} as in Ref. \cite{bnoas_ours}, using GGA for the semi-local exchange-correlation effects. In DFT+U, Hund's coupling was fixed at $J=0.7$ eV. For the magnetic calculations on BNOAS, the on-site Coulomb repulsion $U$=$7$ eV was chosen for most of the studies. We did check that results are fairly insensitive to somewhat smaller values of $U$. For example, at $U=5$ eV, the hopping parameters differed by at most 10 meV. The energy gap (which is not the Mott-Hubbard gap, see following sections) is small in this range of $U$, closing around $U_c\approx$5.5 eV (specifically, $U_c$= 5.2 eV and 5.8 eV in {\sc fplo} (Full Potential Local Orbital)\cite{fplo} and {\sc wien2k}, respectively). We note that in {\sc wien2k} $U$ and $J$ are applied within an atomic sphere. The sphere does not include quite all of the atomic orbital (whose tails are not well defined anyway), also the atomic function is distinct from the Wannier function (see below) that might be argued to be a more realistic `local orbital'. Both of these aspects would require, and perhaps justify, a somewhat smaller value of `atomic' $U$ to achieve the same impact of on-site repulsion.

To construct the maximally localized Wannier functions, we used the {\sc wannier90} \cite{wan90} and {\sc wien2wannier} \cite{wien2wan} codes, projecting onto symmetry adapted Wannier functions.
Two frozen energy windows were investigated: --5.7 eV to 1.3 eV (--7 eV to 3.5 eV) for 27 (37) starting orbitals, given below, for nonmagnetic (respectively magnetic, textured singlet) BNOAS.
The parameters for BNOAS, included in Table I, are obtained for Ni $d$, O $p$, Se $(s),p$, and Ag $(s,p),d$ orbitals (the orbitals inside parentheses are the additional orbitals used for the singlet state).

The excellent quality of the Wannier bands for NM BNOAS are shown in Fig. \ref{nmwan}; the corresponding result for magnetic BNOAS is displayed in Appendix A.
For comparison, we also calculated Wannier parameters for NM LaNiO$_2$. For a fit with more flexibility compared with the previous results of Botana and Norman \cite{botana2020} where Ni $d$, O $p$, and La only $d_{z^2}$ orbitals were used, we used 23 starting orbitals of La $d,f$, Ni $d$, and O $p$ orbitals within an energy widow of --8.5 eV to 1.8 eV. The excellent representation of the bands is shown in Appendix A.

 \begin{figure}[tbp]
  \includegraphics[width=0.85\columnwidth]{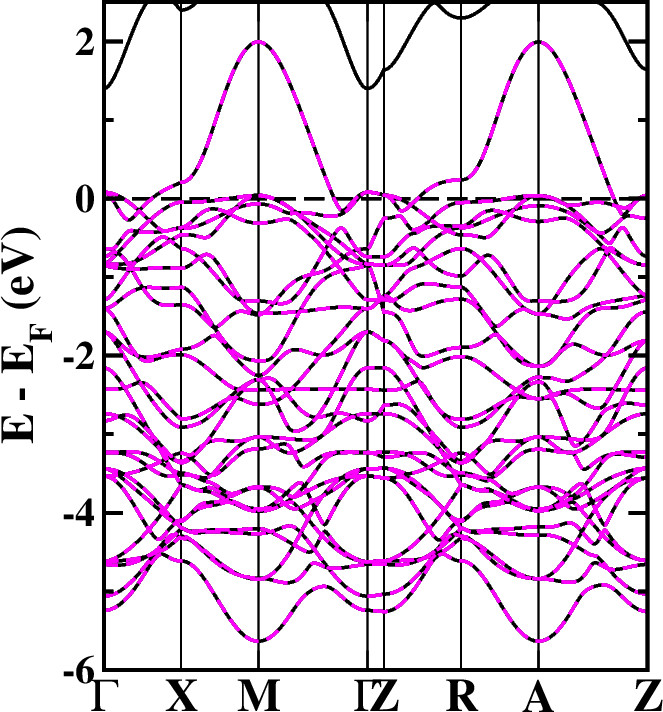}
  \caption{A full valence band Wannier fit to the band structure for NM BNOAS; the two are indistinguishable. As an initial starting point, the following 27 orbitals were used: $d$ orbitals of Ni and Ag ions, and $p$ orbitals of O and Se ions.
  }
  \label{nmwan}
\end{figure}

\section{Electronic and Magnetic Structure}

\begin{table}[bt]
\caption{On-site energies and hopping parameters (in units of eV),
obtained from the transformation to Wannier functions using 27 (37) orbitals in the unit cell (see text),
for NM ($U=0$) and singlet ($U=7$ eV) states of BNOAS. 
NM \lno~ is provided for comparison.
The O1 [$\frac{1}{2}$,0,0] and O2 [0,$\frac{1}{2}$,0] ions sit along the $\hat{a}$- and $\hat{b}$-direction, respectively.
Only near-neighbor hoppings are listed, others for further neighbors are much smaller although necessary for excellent fits.
Boldface and italics denote values for specific attention.
Here, $\Delta_{cf}$, $\delta_{cf}$, and $\delta^p_{cf}$ indicate the crystal field splittings between the centers of $t_{2g}$ and $e_g$ orbitals, between $e_g$ orbitals, and between $p_\sigma$ and $p_\pi$ orbitals, respectively.  
}
\begin{center}
\begin{tabular}{lrrrr}\hline\hline
 ~~On-site energies~~& ~~\lno~~ &\multicolumn{3}{c}{BNOAS}\\\cline{3-5}
 ~~&                 ~~NM~~ & ~~NM~~ & \multicolumn{2}{c}{Singlet}\\
  ~ & ~& ~& ~~up~~ & ~~dn~~\\\hline
 $d_{xy}$ & -1.62 & -1.10 & -3.84 & -4.36 \\
 $d_{xz, yz}$ & -1.46 & -0.79 &-4.57 & -4.24 \\
 $d_{x^2-y^2}$ & {\it -1.04} & {\it -0.74} & {\bf -5.27} & {\bf  2.62} \\
 $d_{z^2}$     & {\it -1.45} & {\it -0.67} & {\bf  2.43} & {\bf -5.61} \\
 $\Delta_{cf}$  & 0.27       &     0.19    &             &             \\
 $\delta_{cf}$  & 0.41       &    -0.07    &             &             \\
 
 \hline
 $p_x=p_{\sigma}$ O1 & -4.85 & -3.27 & -2.56 & -2.38\\
 $p_y=p_{\pi}   $ O1 & -3.34 & -2.06 & -1.97 & -2.01 \\
 $p_z$            O1 & -3.39 & -2.09 & -1.49 & -1.49 \\
 $p_x=p_{\pi}   $ O2 & -3.34 & -2.06 & -1.97 & -2.01 \\
 $p_y=p_{\sigma}$ O2 & -4.85 & -3.27 & -2.56 & -2.38 \\
 $p_z$            O2 & -3.39 & -2.09 &  -1.49 & -1.49\\
 $\delta^p_{cf}$& 1.51       &    1.21    &             &             \\
 \hline  \hline
~~Hopping\\ ~~amplitudes~~& ~~ & ~~ & ~~ & ~~\\
\hline
$d_{xy}-p_{\pi}$    O1 & {\bf -0.65} & {\bf -0.53} & -0.48 & -0.47 \\
$d_{xz}-p_z$        O1 &      -0.70  &       0.56  &  0.48 &  0.48\\
$d_{x^2-y^2}-p_{\sigma}$ O1 & {\it 1.25} & {\it 0.97} & 0.98 & 1.03 \\
$d_{z^2}-p_{\sigma}$     O1 & -0.46       & {\bf -0.28} & {\bf -0.70} & {\bf -0.68}\\
$p_y$ O2 $- p_x$ O1 &  {\it  -0.44} & {\it -0.51} & -0.08 & -0.09\\
$p_x$ O2 $- p_x$ O1 & 0.16 & -0.22 & -0.19 & -0.20\\
$p_y$ O2 $- p_y$ O1 &  0.16 & -0.22 & -0.19 & -0.20\\
$p_x$ O2 $- p_y$ O1 & -0.23 & -0.11 & -0.13 & -0.13 \\
$p_z$ O2 $- p_z$ O1 & -0.15 & -0.01 & 0.07 & 0.09\\\hline\hline
\end{tabular}
\end{center}
\label{table1}
\end{table}

\subsection{$d^8$ versus $d^9$ configuration}
Many of the aspects of the electronic and magnetic structures of BNOAS were presented in our earlier paper \cite{bnoas_ours}, and referring to Fig. 4 of that paper will facilitate the reading of this section.
A different perspective is gained from identifying how the BNOAS and ${\cal R}$NiO$_2$ electronic structures of the NiO$_2$ electronic
parameters differ in the  Ni $d$ - O $p$ energy range, given the same infinite layer scheme (no apical oxygen) with somewhat differing environments. 
The on-site energies and hopping amplitudes obtained from Wannier function representations are presented in Table~\ref{table1} for ${\cal R}$=La. 
Also shown are the differences between NM and the magnetic BNOAS parameters, to indicate the changes that result from applying $U$=7 eV for correlation effects.

\vskip 3mm
{\it Orbital site energies: nonmagnetic case.}
On the broadest level, one can consider the mean position of all five $d$ levels relative to the Fermi level $E_F$, something that is not often considered because the $t_{2g}$ states are so strongly bound and inert. The mean $3d$ site energies are -0.82 in BNOAS, -1.41 in \lno, indicating an overall ``chemical shift'' upward of  0.6 eV in BNOAS of the Ni site. This shift will be due largely to the difference of nearby cation, Ba$^{2+}$ in BNOAS versus La$^{3+}$ in \lno, with some difference arising the different formal charge states. 

The $e_g$ crystal subsplitting, $\delta_{cf}=\varepsilon_{x^2-y^2}-\varepsilon_{z^2}$, 
is reduced drastically and changed in sign in BNOAS (-0.07 eV) compared to 0.41 eV in \lno.
This \lno~ character is common in cuprates, leading to many considerations as a single-band system at zero-th order.  It is highly unusual that $d_{z^2}$ should lie at or above the $d_{x^2-y^2}$ energy, although the difference in site energies in BNOAS is nearly negligible.  These differences will assume importance in later analysis, especially the effective degeneracy of the two $e_g$ orbitals. However, bandwidths (next section) need also to be considered.  

The charge transfer (CT) energy $\varepsilon_d-\varepsilon_p$ receives close attention in transition metal oxides. From the mean Ni $d$ and O $p$ site energies, the CT energies are 2.45 eV for \lno, 1.65 eV for BNOAS. 
This difference leaves BNOAS nearer the CT insulator -- Mott-Hubbard (MH) insulator crossover, in between the MH insulator nickelates and the CT insulator cuprates.
\lno ~itself is usually considered as closer or in fact in the MH insulator regime \cite{berciu2020}. Of course, the ${\cal R}$NiO$_2$ infinite layer nickelates have pockets of carriers of other character \cite{lee2004,choi2020}, thus are considered by many as self-doped Mott insulators \cite{ZYZ2020,lechermann2020,karp2020}.

\vskip 3mm
{\it Hopping amplitudes.} The $d$-$p$ hopping amplitudes are of special interest since, with the site energies, they determine band positions and widths and are of interest in evaluating exchange couplings. Since the $p$ bands are fully occupied, the differences in $p$-$p$ hopping are not of interest here, but are included in Table~\ref{table1} for reference.   
As in cuprates, the in-plane hopping $t_{dp\sigma}$ is largest ($\sim 1$ eV), but not necessarily dominant. Some $d$-$p$ hoppings are quite different between the two compounds. 
The hopping $d_{x^2-y^2}$-$p_{\sigma}$ = $t_{dp\pi}$ differs by 25\%: 1.25 (\lno) versus 0.97 (BNOAS). Other $d$-$p$ hoppings differ by less. These differences affect the shape and the bandwidths of the important bands, which have $e_g$ character. One interesting feature is that magnetism increases the $d_{z^2}-p_{\sigma}$ hopping amplitude by a factor of 2.5.

We note that some site energies and hopping amplitudes are sensitive to what orbitals are included in the Wannier basis set and the energy window. This sensitivity should be taken into account when comparing these quantities between different practitioners.

\subsection{DFT band structure}
The tight-binding parameters discussed in the previous section can give an incomplete picture of the electronic structure. The NM bands of BNOAS are shown in Fig.~\ref{nmwan}. Whereas the two $e_g$ on-site energies are essentially the same ($d_{z^2}$ is slightly higher), only the $d_{x^2-y^2}$ band extends above E$_F$. It does so by 2 eV, due to the usual large hopping $t_{dp\sigma} \approx$1 eV. The $d_{z^2}$ bandwidth arises from twice hopping $d_{z^2}$-$p_x$-$d_{z^2}$ with small amplitude, which (with other mixtures) leaves the $d_{z^2}$ orbital roughly doubly occupied and $d_{x^2-y^2}$ roughly doubly empty. \lno~ is much like cuprates, with its $d_{x^2-y^2}$ orbital about half-filled (with complications due to the La $d$ orbital dispersion).

 \begin{figure}[tbp]
  \includegraphics[width=\columnwidth]{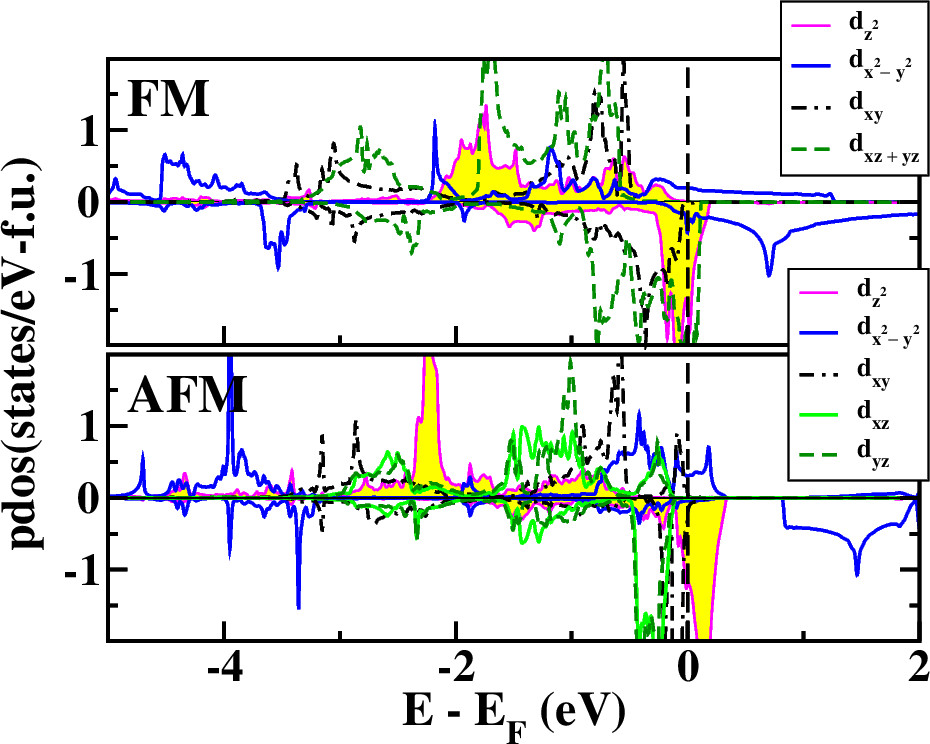}
  \caption{Ni $d$ orbital-projected densities of states (DOSs) in (top) FM and (bottom) AFM at the DFT level. These states are $S=1$ states, {\it i.e.} magnetic Ni. Minority DOSs are plotted downward.
  }
  \label{magBNOAS}
\end{figure}

In the FM state, the total moment in GGA is 1.29 $\mu_B$/f.u., consisting of atomic (sphere) moments of 1.04 for Ni, 0.07 for O, and 0.03 for Se (in units of $\mu_B$). This reduction from 2$\mu_B$ reflects itinerant character and also indicates significant $p-d$ hybridization accounting for reduction from a naive Hund's value. 
The AFM state has a nearly identical Ni local moment.

Considering spin-ordering, only $S$=$1$ states ({\it i.e.} magnetic Ni) are obtained at the DFT level.
Figure \ref{magBNOAS} shows the Ni $d$ orbital-projected DOSs for the AFM and FM alignments.
In the minority channel, the $d_{x^2-y^2}$ band is fully unfilled and the $d_{z^2}$ band is partially empty, with the other bands fully filled. Note that the minority $d_{z^2}$ band is well localized around $E_F$ (narrow peak) in both FM and AFM alignments. 
In the majority channel, most $d$ bands are nearly filled, but some of the dispersive $d_{x^2-y^2}$ band crosses above $E_F$.

\begin{figure*}[!ht]
  \includegraphics[width=1.8\columnwidth]{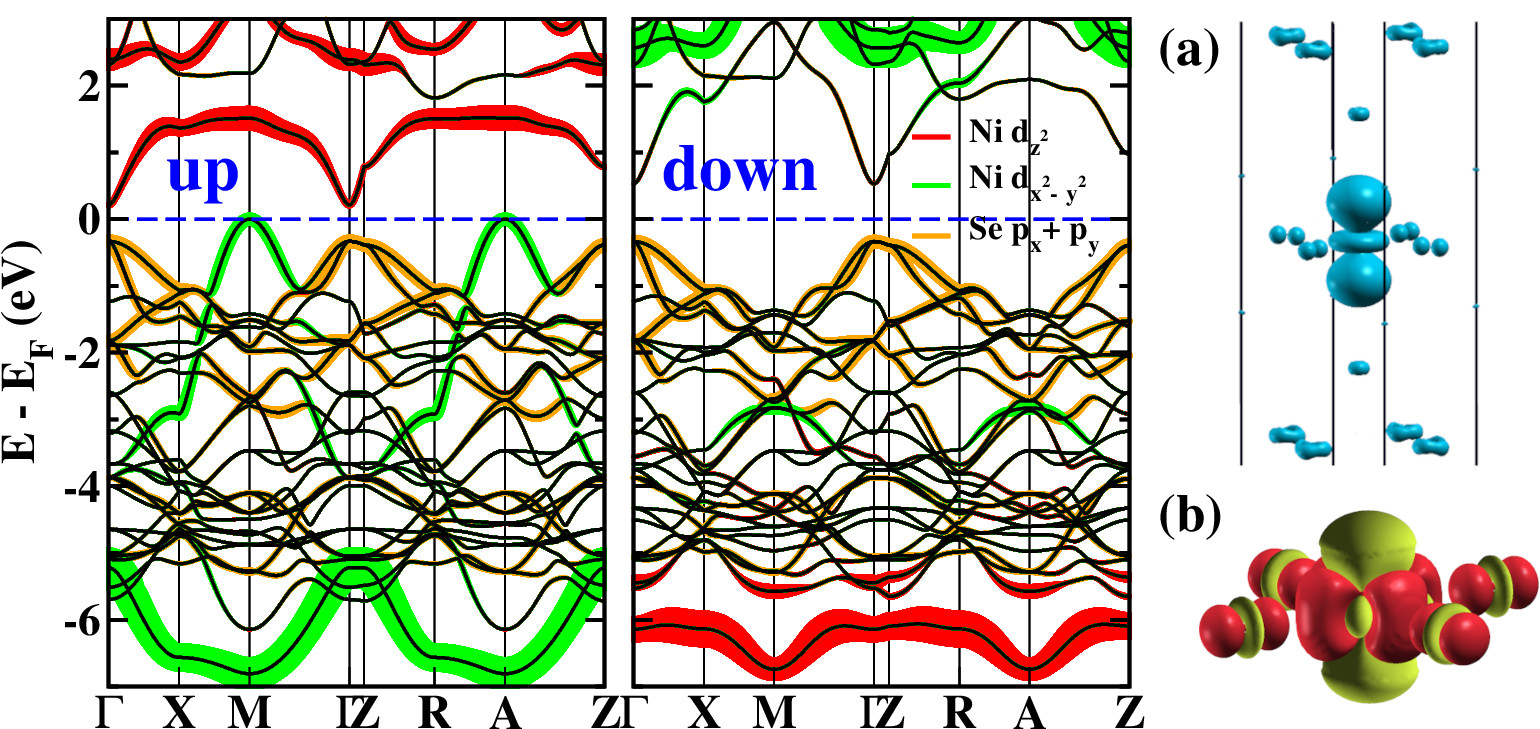}
  \caption{Left panel: The spin up and spin down bands of the singlet band structure at $U=7$ eV. Colors and their thickness of the fatbands indicate the degree of orbital character as denoted by the inset. Right Panels: (a) Hole spin density isosurface plot containing the bottom of the conduction bands in the range of 0.2 eV to 1.7 eV. In addition to the characters on the NiO$_2$ layers, Se $p_x+p_y$ and Ag $s$ characters are visible.  
   (b) Spin density isosurface plot, which is the integral over the valence bands of the difference of up and down spin densities. The oxygen participation in the CuO$_4$ unit is evident (see text). In the isosurface plots, the isovalue is 0.015 $e$/\AA$^3$.
  }
  \label{ods}
\end{figure*}

\begin{figure*}[!ht]
 \hskip -7mm
  \includegraphics[width=1.7\columnwidth]{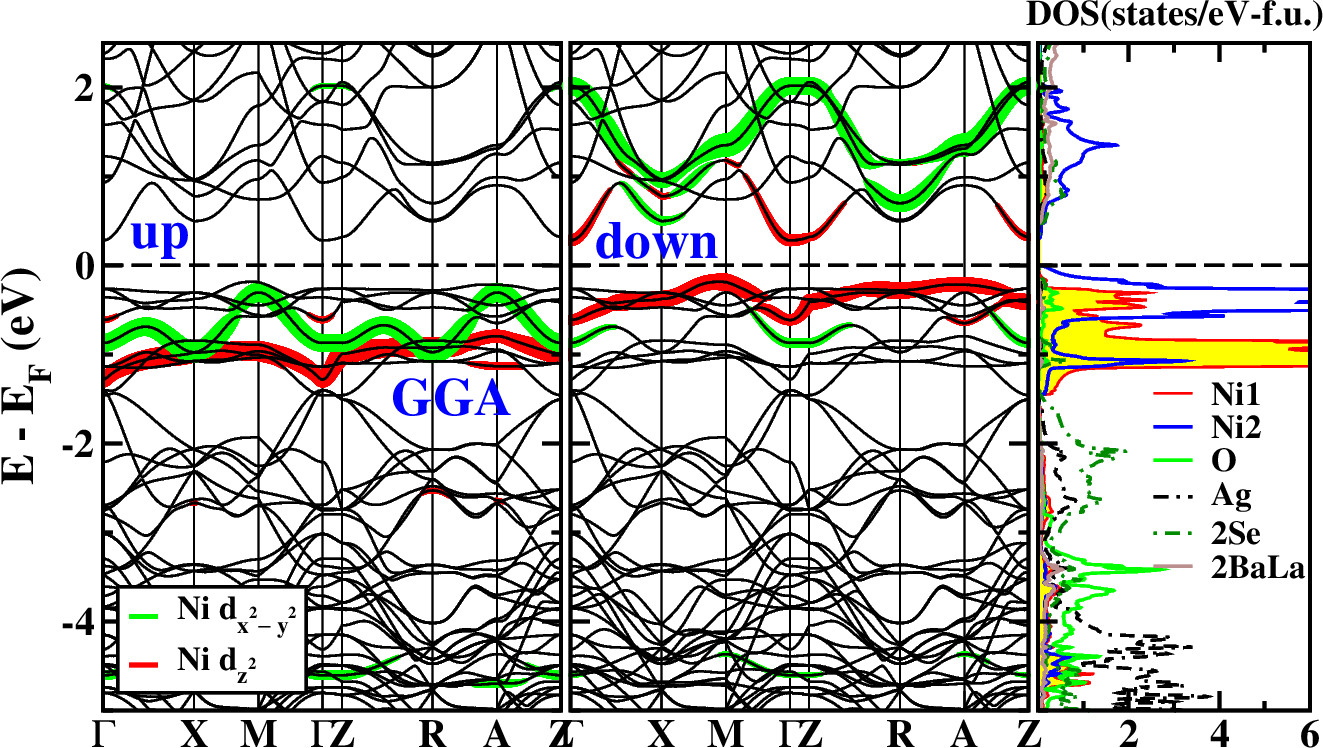}
  \caption{Virtual crystal band structures and atom-projected DOSs of AFM ($X^{2.5+}$)$_2$NiO$_2$(AgSe)$_2$ at the DFT level, showing an energy gap of 0.3 eV. The Ni configuration is $d^9$, with a minority $d_{x^2-y^2}$ hole. }
  \label{dopingAFM}
\end{figure*}

\section{Toward a $d^9$ state}
\subsection{DFT+$U$ electronic structure}
The correlated DFT+U band structures lead, as anticipated, to magnetic Ni ions, with the bands shown in Fig.~\ref{ods} (left panel).  In this calculation all Ni atoms have the same atomic configuration -- the unit cell is not doubled. At first look, the spin-up versus spin-down differences seem minor. This state occurs even when applying a small $U$ such as $U_{eff}$=$U$-$J$$\approx0.3$ eV with $J=0.7$ eV in the DFT+U approach \cite{bnoas_ours}.

Further analysis reveals qualitative differences: there is spin polarization, but the number of occupied up (or down) bands is the same, so the net moment vanishes (the choice of `up' [majority $d_{x^2-y^2}$] and `down' was arbitrary).  Yet the system has FM rather than AFM symmetry; the latter has natural symmetry dictated vanishing net moment in a doubled cell, and identical up and down bands. Correlation has produced in BNOAS a state with the symmetry of a FM -- distinct up and down bands but no broken space group symmetry -- but vanishing moment. This result is indicative of an {\it anti-Hund's magnetic state}. 
Note that inclusion of $U$ also has increased the moment to near the ideal Hund's high-spin value of 2$\mu_B$/Ni for local-moment AFM or FM order.

This symmetry analysis is reminiscent of a so-called half metallic AFM (or compensated half-metal) \cite{wep01,wep08} but is insulating. What has occurred is that Mott insulating behavior has emerged in both $d_{x^2-y^2}$ and $d_{z^2}$ orbitals, but with spin-reversed occupation (or equivalently, unoccupied) orbitals. The spin-half configurations of each orbital are coupled to a singlet. The issue of classical (band) singlet versus quantum (correlated orbital) difference is discussed in Secs. IV and V.

Due to other active orbitals in the gap region, the band gap is not the MH gap. The gap-edge states involve Ba $d$, Se $s,p$, and Ag $s$ orbitals, and these bands become somewhat magnetically polarized, differently for the up and down spin directions. Similar bands in ${\cal R}$NiO$_2$ materials overlap E$_F$ and figure strongly in theoretical models.\cite{botana2020,berciu2020,choi2020,lechermann2020,karp2020}

The next step is to consider orbital character of bands more closely, and to notice that for the up bands, and unlike cuprates, the unoccupied $d$ band unexpectedly has $d_{z^2}$ character, while the down spin hole is in the $d_{x^2-y^2}$ orbital.  This canceling of spin-half moments in different orbitals was denoted previously \cite{bnoas_ours} as an ``off-diagonal singlet'', off-diagonal in orbital giving it an anisotropic spin density with vanishing net moment, yet having exchange coupling to neighbors through the Ni $d_{x^2-y^2}$ orbitals. Our original report \cite{bnoas_ours} provided more description of the electronic structure. This DFT+$U$ singlet is, to our knowledge, unique in the condensed matter physics literature, although such correlated magnetic states have long been studied by quantum chemists \cite{qchem}.

\subsection{Virtual crystal electron doping}
To investigate a $d^9$ system like ${\cal R}$NiO$_2$, one Ba ion is replaced by a La ion.  Using the virtual crystal approximation, both Ba ions are replaced by a virtual ion $X^{2.5+}$ with nuclear charge Z=56.5, restoring the lattice symmetry and adding one electron/f.u. which goes to the Ni ion.
In this calculation, we used the lattice parameters of BNOAS \cite{bnoas_ours}, but the internal parameters are optimized with {\sc wien2k}.

At the DFT level, the AFM state is energetically favored over FM and NM states by 116 and 155 (in units of meV/Ni), respectively.
In both spin-ordered states, the Ni local moment is 0.84 $\mu_B$, consistent with a strong $S=\frac{1}{2}$ spin reduced somewhat by hybridization, as is common. 
These energy differences lead, in a fixed moment nearest neighbor Heisenberg picture, to a superexchange $J=116$ meV.  Comparing with values for infinite-layer nickelates $J$=64 meV from spin wave theory \cite{leonov2020} and spectroscopy \cite{Lu2021}, and the range of values reported for cuprates $J$=125-158 meV \cite{cuprate2017j}, this value for La-substituted BNOAS lies in an intermediate range. The similarities suggest that electron-doped BNOAS is a candidate for magnetically coupled superconductivity, perhaps with higher T$_c$ than known nickelate superconductors.

The AFM band structure is shown in Fig.~\ref{dopingAFM}. Since a gap opens already for $U$=0, we show the results at the GGA level.
At this virtual crystal level, doped holes go primarily into Se $s,p$ bands along the $M$-$A$ line. This prospect is much different than discussions of cuprates and related transition metal oxides, where the question is MH insulator versus CT insulator and where the respective holes reside. In BNOAS the blocking layer is becoming active, receiving both electron- and hole-doped carriers. Prospects for superconductivity are a topic for the future, but it seems clear that doped BNOAS will be distinct in behavior from doped cuprates and most transition metal oxides, with some potential similarity to infinite-layer nickelates.

\section{Spin polarization;\\ the anti-Hund's singlet} 
The character of doped holes and their relationship to superconductivity in HTS cuprates have been a central concern. The CT, versus MH, characterization of the energy gap leads to a large amount of Zhang-Rice (ZR) singlet character \cite{zhangrice} in doped holes and the electron removal spectrum. The ZR singlet is a coupling of spin-half Cu spin with an oppositely directed linear combination of planar O $p_{\sigma}$ orbitals of $d_{x^2-y^2}$ symmetry. Holes in a Mott insulator would have simple $d_{x^2-y^2}$ character, with small (negligible) O character.
The $d^8$ Ni ion in BNOAS presents a different reference state, yet singlet versus triplet character becomes a central feature, albeit on-site rather than involving oxygen.

Including $U$ in the electronic structure calculation to account for
on-site repulsion in a mean field picture, and thereby obtaining magnetic character, causes important changes in the electronic structure and the derived band parameters, as shown in Table~\ref{table1} under the two rightmost BNOAS columns. First, it should be expected that all Ni $d$ orbitals will be shifted: strongly downward if occupied, somewhat upward if unoccupied. These differences will appear most apparently in the on-site parameters, but a secondary but still significant effect can be seen in Wannier hopping amplitudes. 

BNOAS presents in the calculated ground state not the trivial singlet of a completely unoccupied $d_{x^2-y^2}$ orbital, but a singlet with internal character as described in the previous section: there is exchange coupling within the NiO$_2$ layer and weaker coupling between layers, leading to the Kondo sieve model \cite{bnoas_ours} for the insulating spin system. The results that we present for BNOAS are for this singlet phase, which persists in our calculations for $U>4$ eV, using two different all-electron DFT codes {\sc wien2k}\cite{wien2k} and {\sc fplo}\cite{fplo} and with various starting spin configurations. In BNOAS, for each spin direction all five majority orbitals are occupied while one minority orbital is unoccupied. This peculiar singlet that we find to be persistent differs in which $e_g$ spin-orbital is unoccupied. Because the spin splitting between majority and minority is so large, the mean $d$ site energy in the magnetic state becomes less relevant.

 For the Ni $t_{2g}$ orbitals, the in-plane $d_{xy}$ orbital is more strongly bound than the other two by 0.3-0.4 eV, an effect of environment and bonding-type hybridization in-plane. The $e_g$ orbitals are of primary interest. The mean $e_g$ on-site energies (majority, minority; see Table I) 
 are nearly degenerate. Individually, the exchange (up minus down) site energies are around 8 eV, equal to $U$=7 eV plus some contribution from Hund's $J$ and hybridization effects.   The mean $d$ sites energies are 1.0-1.3 eV lower for the singlet case, because there are more occupied majority orbitals (moved down) than minority orbitals (moved up). Regarding oxygen, only the $p_{\sigma}$ orbitals show a significant spin dependence, with majority being 0.2 eV lower than minority.  

For the hopping amplitudes of BNOAS, only one shows a significant difference due to singlet formation relative to the NM state. This one is the $d_{z^2}$-$p_{\sigma}$ hopping, normally considered to be insignificant in square-planar geometry. While it is -0.28 eV in NM BNOAS, in the singlet case it becomes -0.7 eV for both spins. The effect is to spread $d_{z^2}$ character into band positions (bonding and antibonding) where it would otherwise be negligible. 

{\it Origin of the Singlet.} The underlying feature of significance of the GGA band parameters for anti-Hund's singlet formation is the effective degeneracy of the two $e_g$ orbital energies. One could consider a unitary transformation among the $e_g$ orbitals but reducing the symmetry without any evident usefulness does not appear attractive. The anti-Hund's orientation for spin-half moments costs $J_H/2$ = 0.35 eV in exchange energy, for the value of $J_H\approx J$=0.70 eV we have used in our DFT+$U$ calculations (see Sec. II). The value of pair spin $S$=0 or $S$=1 ({\it i.e.} the relative spin density orientation) affects the band structure and hence the kinetic energy, the charge and spin densities, and the correlation energy, with some contribution from the purely Coulombic (charge density) terms. The energy sources favoring the anti-Hund's orientation are not clear. 
It is worth noting that in \lno, where the $e_g$ subsplitting is of the usual sign with a value of 0.7 eV, even then in the large $U$ regime a similar singlet character (opposite orientation of $e_g$ spins) became dominant \cite{lee2004}.  
Studies varying the lattice constant, hence the interatomic and interionic separations, could be helpful in understanding the mechanism of singlet formation.

\section{A quantum singlet in an itinerant background}
\subsection{A minimally correlated wavefunction} 
The Kohn-Sham (KS) treatment within DFT provides an effective single-particle (mean field) picture of the system. The `singlet' discussed so far is the classical-spin version of one spin up, one spin down on each Ni ion, with no quantum nature of the state. A generalization to a quantum singlet in an unpolarized background (at least in the atomic limit) can be represented by a  wavefunction that is a generalization of the KS or Hartree-Fock (HF) one:
\begin{equation}
    \Psi(\{r_{i,s}\}) = {\cal A} ~~ Q(r_{1,\uparrow},r_{2,\downarrow})
        D^{\uparrow}(\{r_{j,\uparrow}\}) D^{\downarrow}(\{r_{j,\downarrow}\}),
        \label{wavefn}
\end{equation}
Here $Q$ is the singlet wavefunction for particles 1 and 2, and $j$ runs only over the other particle indices. $D^{\uparrow}$ ($D^{\downarrow}$) is the Slater determinant for the occupied particle states (except
 for singlet orbitals) with spin up (down). ${\cal A}$ is the antisymmetrization
 operator which finally provides a function $\Psi$ that is antisymmetric under exchange of any two particles' coordinates.  This wavefunction is heuristic, as the issue of projecting out particle coordinates 1 and 2 out of the determinants is a remaining problem. (Also, in actual application it would contain unscreened exchange and no correlation.)

 The determinants contribute nothing 
to the net spin polarization unless the orbitals are spin-dependent.
The HF states, for local orbitals $\phi_1, \phi_2$ involved in a singlet ($e_g
$ orbitals for BNOAS), are $\phi_m(1,s)=\phi_m(1)|s\rangle$ in terms of the spin
or $|s\rangle$ ($s=\pm 1$ for up and down respectively) and the argument of $\phi_m$ is simplified to $\phi_m(1)$ for particle 1, etc.
and the orbitals are the same for both spins (by antisymmetry).
The spin-singlet wavefunction is
\begin{eqnarray}
    Q(1,2)&=&\frac{1}{\sqrt{2}}([\phi_1(1)\phi_2(2)+\phi_1(2)\phi_2(1)] \nonumber \\
     & &\times \frac{1}{\sqrt{2}}[ |\uparrow\downarrow\rangle-|\downarrow\uparrow\rangle]
\label{singlet}
\end{eqnarray}
in common notation.

\subsection{Features of the wavefunction}
The charge (+) and spin (-) density operators are
\begin{equation}
    {\hat n}_{\pm}(r) = \sum_i [\delta(r-r_{i,\uparrow}) \pm \delta(r-r_{i,\downarrow})].
\end{equation}
Taking the matrix elements in $|Q\rangle$, one obtains that $n_+(r)$=$\phi^2_1(r) + \phi^2_2(r)$ (which are the KS or HF values) and, from spin inversion of $Q$ giving only a change in sign, the spin density $n_{-}(r)$=$m(r)$ is identically zero. Thus singlet formation does not change the total density, and the spin density vanishes (at this level of description) even for anisotropic orbitals $\phi_m$: the up amplitude of each orbital cancels its down amplitude. Notice however from Eq.~(\ref{singlet}) that the singlet wavefunction itself is textured, {\it i.e.} anisotropic, allowing exchange coupling between sites. If the quantum nature of the singlet is quenched (as in DFT theory) the non-zero spin polarization is of an incomplete singlet type, similar to that seen earlier \cite{lee2004}.

The correlated HF-singlet wavefunction of Eq.~(\ref{singlet}) brings new terms into the hybrid Kohn-Sham HF equations. The kinetic energy for a given particle includes new `spin exchange' terms
\begin{equation}
\nabla^2 \Psi = (\nabla^2 Q){\cal D} + Q(\nabla^2{\cal D}) +2\vec{\nabla} Q\cdot\vec{\nabla}{\cal D}
\end{equation}
where ${\cal D}$ denotes the product of the two Slater determinants. The singlet couples the spin directions, bringing in new `exchange' energies from $Q$ in the kinetic energy. Some alteration in the usual Coulomb exchange potential will also arise from the singlet spin-mixing. Further analysis requires more specificity of viewpoint, due to the substantial differences in intent and in form of the HF and the KS equations, or a possible hybrid HF-KS treatment.

\subsection{Excitations above the singlet phase} 
BNOAS presents interesting complications. Considering the correlated HF wavefunction posited above, the actual existence of the singlet ground state presents questions. With the spin-coupled $e_g$ orbitals removed from the band structure (the Slater determinants) the calculated band structure (with compensating classical spins) becomes approximate.   In addition, one of the orbitals, say $\phi_1$ ($d_{x^2-y^2}$) is coupled to neighboring $\phi_1$ orbitals in-plane, favoring antialignment, while the $\phi_1$ and $\phi_2$ orbitals are coupled to a singlet. Neglecting further interactions (discussed by us earlier \cite{bnoas_ours}) this situation is a realization of the Kondo necklace Hamiltonian, in the regime where Kondo coupling is dominant.

For weaker 
 Kondo coupling the $S=1$ state can compete with the $S$=$0$ singlet, with some results in this regime having been explored by Singh \cite{RRPSingh}.  The triplet nature of the ion, with orientations $M_S$=+1,0,-1, and anticipated local anisotropy due to the environment, provides a rich palette for study. Considering the unusual susceptibility (reviewed in the Introduction) a favored $M_S$=0 orientation may be of special interest to consider for BNOAS.
With a small (calculated) bandgap, band excitations are available to participate in optical conductivity and, virtually, in many-body processes that impact low energy phenomena, including superconductivity.

\section{Discussion and Summary}
Spin singlet physics in 2D systems has played a large role in cuprate HTS, and similarities of nickelates to cuprates have encouraged extension of this concept to superconducting nickelates. The dominant concept has been that of ZR singlet \cite{zhangrice}, building on the characterization of undoped cuprates as CT insulators rather than explicit MH insulators. The latter would involve only $d$ orbital reoccupation by doping or low energy excitations, while in CT insulators doped holes will lie primarily on neighboring O $p$ orbitals. A linear combination on four O sites will form a molecular orbital of ($d_{x^2-y^2}$) symmetry, and a hole in this molecular orbital will pair with the central spin-half Cu$^{2+}$ to form a spin singlet, the ZR singlet. While most studies have assigned nickelates as having more Mott than CT character, the ZR singlet concept is sprinkled through the recent nickelate literature. The infinite layer nickelates are more often considered from a Hund-Hubbard viewpoint \cite{chang2019}.  We find that electronic structure parameters suggest that BNOAS lies in the vicinity of a charge transfer to Mott-Hubbard (CT-MH) crossover, making the basic `physics' harder to pinpoint.

Zhang, Yang, and Zhang (ZYZ) have proposed \cite{ZYZ2020} a generalized singlet construct. Noting that even undoped NdNiO$_2$ has a small concentration of Nd $5d$ carriers mixing somewhat with Ni $d$ states, they have proposed that ZR-like combinations of these carriers form a local molecular orbital and form a singlet. Each occupied Nd singlet removes a spin from the Ni square lattice, and it is proposed that the few percent provided by Nd inhibits AFM ordering of the Ni spins.  These ZYZ singlets might also account for observed low-T transport behavior. The model was built on the Mott insulator viewpoint before the CT insulator picture of nickelates became more prevalent. Photoemission spectroscopies are now indicating charge or spin ordering \cite{Tam_CDW_2021,Krieger_CDW-SDW_2021,Rossi_CO_2021} in ${\cal R}$NiO$_2$ (${\cal R}$ a lanthanide atom), complicating any connection between such singlets and inhibition of geometrical symmetry breaking.

Singlet phases arise in the 2D Shastry-Sutherland spin model in certain parameter regimes, but they have a different character and origin than those discussed above. On a square lattice of $S$=$\frac{1}{2}$ spins, a specific regime of exchange parameters can result in the formation and ordering of singlets from spins on neighboring sites or across the diagonal of the square \cite{Haravifard2014}. Such exchange processes enrich the phase diagram.

In this paper it has been shown by comparison in what ways electronic structure parameters differ between $d^8$ BNOAS and $d^9$ LaNiO$_2$. Both share the infinite-layer local environment of the Ni ion, but the further environment (Ba$^{2+}$ versus La$^{3+}$ etc.) affects several parameters. Separately, the strongly increased in-plane {\it a} lattice constant, with related perpendicular changes, has a strong influence on the physics.

\section{Acknowledgments} We acknowledge communications with A. S. Botana, M. R. Norman, and V. Pardo, and conversations with R. R. P. Singh, and feedback from Wei-Lin Tu and Hyun-Yong Lee.
 H.S.J. and K.W.L. were supported by National Research Foundation of Korea Grant No. NRF-2019R1A2C1009588. 
W.E.P. was supported by NSF Grant No. DMR 1607139.

\appendix

\begin{figure*}[!ht]
  \includegraphics[width=1.6\columnwidth]{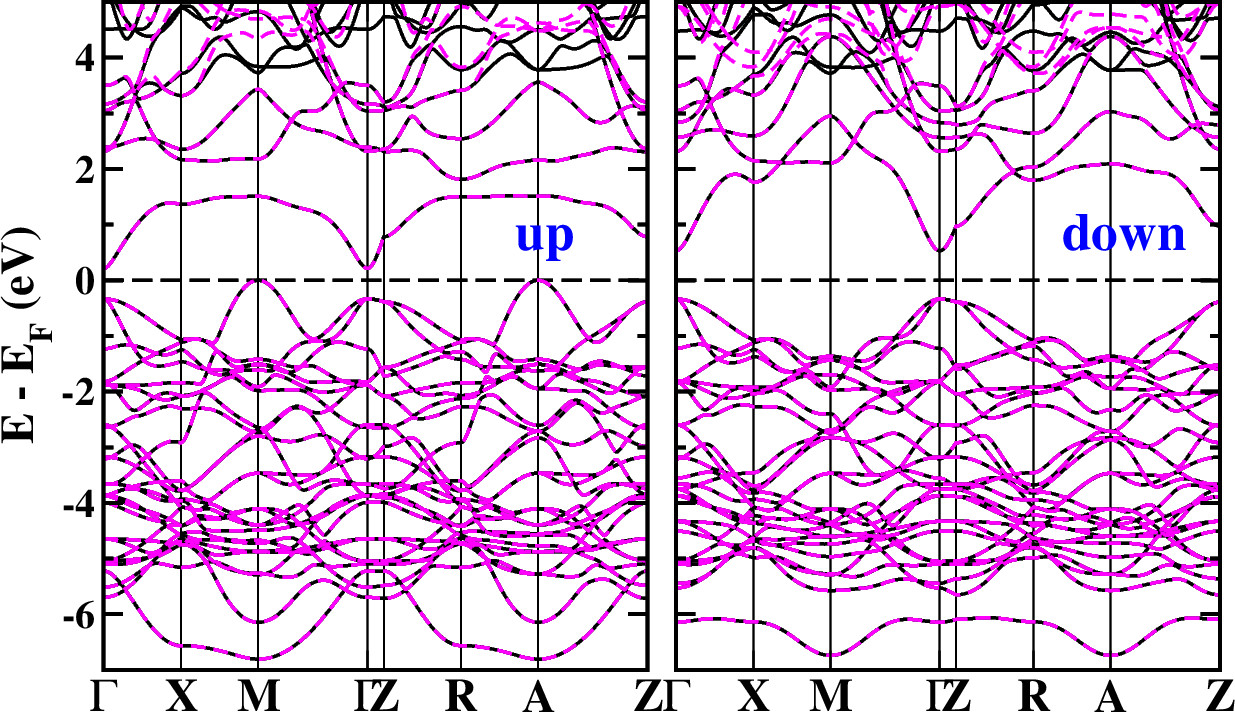}
  \caption{Wannier fitting for the textured singlet phase of BNOAS at $U=7$ eV. Dashed red lines indicate the fit to the calculated bands (solid black lines). As the initial starting point for this fitting used the following 37 orbitals: Ni \{$d$\}, O \{$p$\}, Se \{$s$, $p$\}, and Ag \{$s, p, d$\} orbitals. For this fit the Se $s$ and Ag $s$, $p$ bands were added to include lower conduction band hybridization.
  }
  \label{ODSWan}
\end{figure*}

\begin{figure}[!ht]
  \includegraphics[width=0.80\columnwidth]{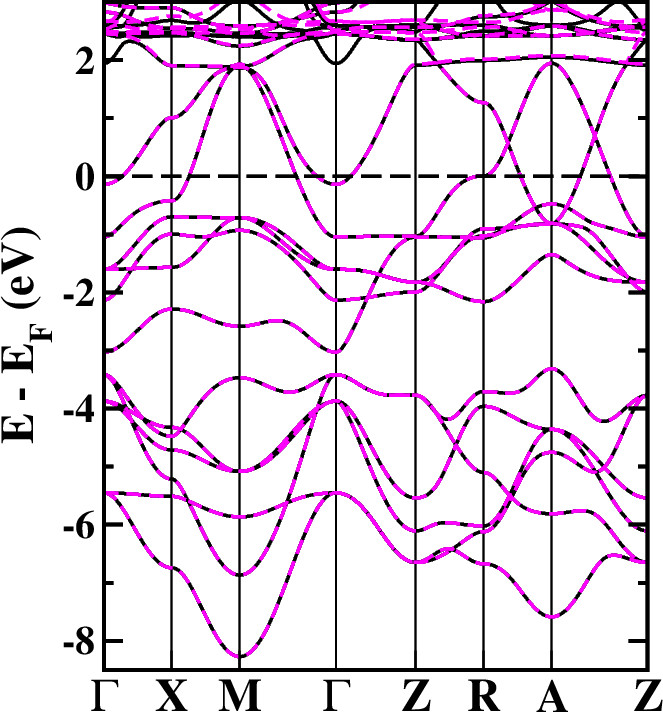}
  \caption{Wannier fit of the bands for NM LaNiO$_2$, providing the parameters given in Table \ref{table1}. For this fit, 23 starting orbitals of La $d,f$, Ni $d$, and O $p$ were used.
  }
  \label{LNOWan}
\end{figure}

\section{Wannier fits to \\ the Textured Singlet Bands}

Figure ~\ref{ODSWan} provides comparison of the band structures of the textured singlet state of BNOAS. The comparison is for the correlated DFT+$U$ bands, spin up and spin down (black lines), and the Wannier representation that includes 37 orbitals in the primitive cell (dashed red lines that mostly overlie the black lines). The Wannier representation is excellent through the valence bands and in the $d$ bands in the conduction bands. The lower conduction bands are not representative of Mott insulating, basically localized orbital, states because they include large dispersion departing from $\Gamma$ toward $Z$. The dispersion arises from mixing with Se and Ag $s$,$p$ states in the same energy region, through Ba and Ni $d_{z^2}$ orbitals. In the spin up bands, the conduction Mott band would be flat around 1.2 eV except for the mixing. For reference and comparison with Fig. 2, the non-magnetic bands and Wannier fit for LaNiO$_2$ are provided in Fig.~\ref{LNOWan}. 

This character is important in considering doping. Upon adding electrons, they will lie in the (AgSe)$_2$ blocking layer, itinerant and without much coupling to the Ni spin. This holds for both spin directions, although the Mott conduction band for spin down lies around 1 eV higher than for the up spin, and is not easily discernible in the figure.

\section{Band structure of non-magnetic $(X^{2.5+}$)$_2$NiO$_2$(AgSe)$_2$}

\begin{figure*}[!ht]
  \includegraphics[width=1.8\columnwidth]{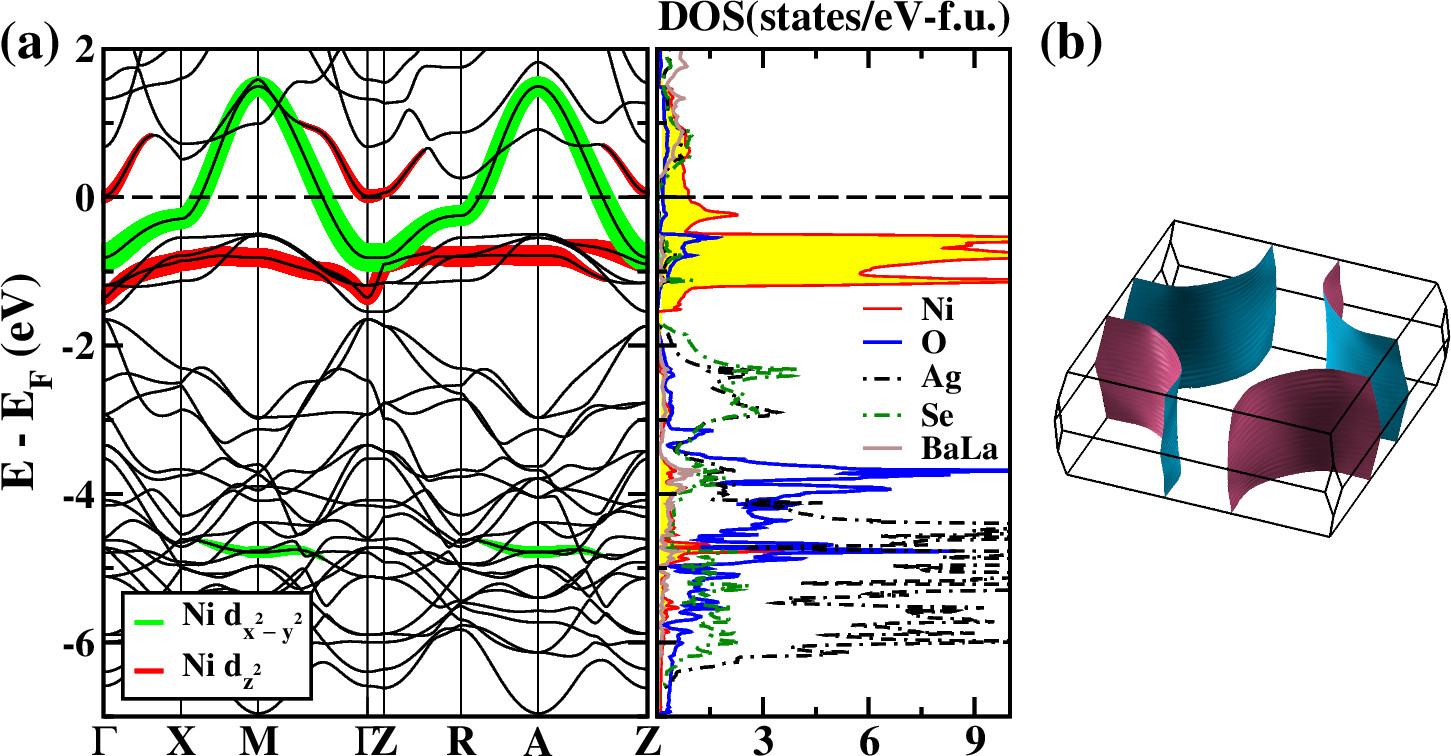}
  \caption{(a) Virtual crystal band structure and atom-projected DOSs, and (b) Fermi surface (FS) of NM ($X^{2.5+}$)$_2$ NiO$_2$(AgSe)$_2$ in the DFT level.   
The FS is a single cylinder centered at the $M$ point.
  }
  \label{dopingNM}
\end{figure*}

Our discussion of electron doping in Sec. IV.B was based on a virtual crystal treatment of the cations in the system Ba$^{2+}$La$^{3+}$NiO$_2$(AgSe)$_2$, leading to ($X^{2.5+}$)$_2$ NiO$_2$(AgSe)$_2$. Figure~\ref{dopingNM}(a) displays the non-magnetic band structure and atom-projected DOS of this system. The DOS can be compared with that of the (preferred energy) AFM DOS in Fig.~\ref{dopingAFM}, and these bands provide a useful reference for how magnetism affects the electronic structure giving the bands of Fig.~\ref{dopingAFM}. The gap lies at the middle of the $d_{x^2-y^2}$ band, leading to the classic $d^9$ half-filled band system in the 2D square lattice. The opening of the AFM gap leads to the narrow $d_{z^2}$ bands lying at the bottom of the gap, which is a central difference compared to $d^9$ cuprate systems, usually discussed in terms of the charge transfer energy difference.

The corresponding Fermi surface, shown in Fig.~\ref{dopingNM}(b), consists of a single 2D cylinder centered at the M point. The circular shape indicates an incipient 2D nesting instability for scattering $|\vec Q|$=2$k_F$ ($k_F$ is the Fermi wavevector) for the half-filled band, contributing to the gap that arises for AFM ordering but not for FM order.


\end{document}